# Urban Geography and Scaling of Contemporary Indian Cities

Anand Sahasranaman[1,*] and Luís M. A. Bettencourt[2,3,#]

[1]Centre for Complexity Science, Dept of Mathematics, Imperial College London, London SW72AZ, UK.

[2]Mansueto Institute for Urban Innovation, Dept of Ecology and Evolution, Dept of Sociology, University of Chicago, Chicago IL 60637, USA.

[3]Santa Fe Institute, Santa Fe NM 87501, USA

[*] Corresponding Author. Email: a.sahasranaman15@imperial.ac.uk

[#] Email: bettencourt@uchicago.edu

**September 23, 2018.**

**Abstract:** This paper attempts to create a first comprehensive analysis of the integrated characteristics of contemporary Indian cities, using scaling and geographic analysis over a set of diverse indicators. We use data at the level of *Urban Agglomerations* in India from the Census 2011 and from a few other sources to characterize patterns of urban population density, infrastructure, urban services, economic performance, crime and innovation. Many of the results are in line with expectations from urban theory and with the behaviour of analogous quantities in other urban systems in both high and middle-income nations. India is a continental scale, fast developing urban system, and consequently there are also a number of interesting exceptions and surprises related to both specific quantities and strong regional patterns of variation. We characterize these patterns in detail for crime and innovation and connect them to the existing literature on their determinants in a specifically Indian context. The paucity of data at the urban level and the absence of official definitions for functional cities in India create a number of limitations and caveats to any present analysis. We discuss these shortcomings and spell out the challenge for a systematic statistical data collection relevant to cities and urban development in India.

## 1. Introduction:

In 2007, for the first time in history, the world population became more urban than rural. This phenomenon is only expected to intensify, with the UN projecting that 66% of the global population will live in cities by 2050, and that over 90% of this increase is expected to be focused in Asia and Africa [1]. Given this context of rapidly increasing urbanization, especially in the developing world, there is an urgency to develop a deeper, scientific understanding of urban processes and their practical implications.

Against this general backdrop of worldwide urbanization, one particular country – India – accounts for the most momentous transformation of all. India is on track to becoming the largest nation in the world by population, expected to surpass China in the next few years with a population around 1.4 billion by 2024 [2]. India's population is expected to peak in the range 1.6-1.8 billion by mid-century. All along, the nation will continue to urbanize from a rate close to 33% today to more than 50% in the same time frame [2]. This translates into approximately an additional 400 million people living in Indian cities in the next three decades, placing pressure on land, urban services, governance structures and general economic development, all of which will need the be substantially transformed very quickly.

The prospect of guiding this massive urbanization successfully in the next few decades demands a much deeper assessment and understanding of Indian cities and their trajectories. Although there is a rich history of case studies, demographic analyses [3–5] and some detailed investigations of specific quantities such as sanitation, slums, and crime [6–10], the present paper – to the best of our knowledge – is the first detailed analysis of Indian cities as complex systems, where a quantitative assessment of many urban attributes is brought together into a common framework.

To do this, we use the framework of urban scaling, together with ideas from urban geography [11–14]. Urban scaling analysis singles out the importance of population size in isolating a set of general agglomeration effects, characterizing economies of scale in infrastructure and urban services, and increasing returns to scale in socioeconomic interactions [11,14]. These effects are the tell-tale signals of cities, and have been observed quantitatively in urban systems from around the world, from the United States to European nations, and from China to South Africa and Brazil [11,12,15–18].

The starting point of the analysis is very simple: Any extensive urban indicator $Y_i(t, N_i)$ for city *i*, with population size $N_i(t)$, at time *t* can be described as:

$$Y_i(t, N_i) = Y_0(t) N_i^{\beta} e^{\xi_i(t)} , \qquad (1)$$

which is an exact expression. The prefactor $Y_0(t)$ is independent of size and carries the physical dimensions of the relevant quantity (e.g. money per year for Gross Domestic Product). Its change in time signifies systemic change for all cities, such as national level economic growth. The scaling exponent $\beta$ is the elasticity of the quantity $Y_i$ relative to population at fixed *t*,

$$\beta = \frac{d \ln Y_i}{d \ln N_i} . \qquad (2)$$

The quantities $\xi_i(t)$ are city specific, scale independent deviations from the scaling law,

$$\xi_i(t) = \ln \frac{Y_i}{Y_0 N_i^{\beta}} \qquad (3)$$

The point about scaling analysis is twofold. First, the exponent $\beta$ is observed empirically to take similar values for broad classes of urban indicators [11,14], with $\beta \simeq 7/6 > 1$, for socioeconomic rates, $\beta \simeq 5/6 < 1$, for spatial density and several kinds of infrastructure, and $\beta \simeq 1$, for household quantities expressing typical individual needs (number of jobs, number of housing units, water consumed at home). Second, these specific exponent values can be computed from urban scaling theory [14], which expresses classical models of urban economics and geography in modern terms, including socioeconomic networks and more realistic transportation costs [19–21].

In this way urban scaling theory – like all earlier mathematical models of urban economics and regional science – defines cities through a budget constraint balancing urban incomes and costs of real estate and transportation [14,19]. This definition translates in practice to so-called functional cities, which are urban areas defined as integrated labour markets or "commute to work" areas; that is regions that comprise together places of residence and work, such as business districts and corresponding residential areas and suburbs. In the United States, this definition corresponds to Metropolitan Statistical Areas [22], which have been constructed by the US Census since the 1950s. In OECD countries these constructions are more recent [23], but have been created at a higher spatial resolution. In India, no similar Metropolitan definition of cities exists at the moment, in part because commuting flows remain hard to measure. Instead, at present, the Census of India builds units of analysis known as *Urban Agglomerations*, which are only an approximation to these concepts. Urban Agglomerations are defined by spatial contiguity, under several additional conditions (see Appendix A). This is an important caveat because urban units of analysis that are "not functional", and are instead defined simply through metrics such as density, or by political boundaries, sometimes are found not to display urban agglomeration properties [24,25].

Despite these caveats, which are at present hard to resolve, we believe that the scaling analysis of Urban Agglomerations is a fundamental first step for a systemic understanding of Indian cities. Below, we explore the scaling relationships of various spatial, socio-economic and infrastructural characteristics of Indian cities with their populations. We analyse the geography of exceptions to average scaling and attempt to make sense of these by invoking a deeper understanding of the geography and history of India. Finally, we discuss how data for Indian cities must improve in the near future and point out to priorities in light of our results and other general considerations from the emerging field of urban sciences.

## 2. Agglomeration and Scaling Effects in Indian Cities

We start by exploring the nature of scaling relationships for three sets of urban characteristics: public infrastructure, social interactions, and individual household needs. We present the detailed discussion on urban units of analysis, data and statistical methods used in Appendix A. We structure our results in two parts - the first deals with the analysis of scaling (agglomeration) effects in Indian cities, while the second pays closer attention to two quantities of great interest, namely crime and technological innovation. Both these quantities show strong regional patterns, which we analyse systematically.

Figure 1 displays a number of scaling diagrams for different public and private types of infrastructure and services.

First, we find that land area shows a behaviour analogous to other urban systems, historically and throughout the world, and with urban scaling theory (expectation $\beta \simeq 5/6$), with an observed sublinear exponent $\beta \simeq 0.88$ (Figure 1B) [26]. This exponent also sets, according to theory, the length of roads as these should be area filling. However, we find instead a somewhat larger, just barely sublinear, value of $\beta \simeq 0.96$ (Figure 1A). It is typical of fast developing nations that formal infrastructure and services are first developed in larger cities and then gradually become more common in smaller places throughout the urban system [15], which may also be the case for India [27]. Other establishments associated with public and social services - such as number (not size) of bank branches and educational institutions - also show sublinear behaviour, roughly tracking land area, suggesting that they occur on average at similar spatial densities in all cities in India big and small, which is again a pattern consistent with other nations. Finally, household services and infrastructure – as measured by private toilets and electricity connections – are roughly proportional (linear) on population, with perhaps just a slightly superlinear effect of a few percent, possibly accounting for the more universal accessibility of these basic services in larger cities [27]. A few cities well below these scaling relations point to major local deficits. Examples are Pondicherry for sanitation, Gorakhpur for banking infrastructure, and Visakhapatnam for education infrastructure, to name a few, see Figure 1.

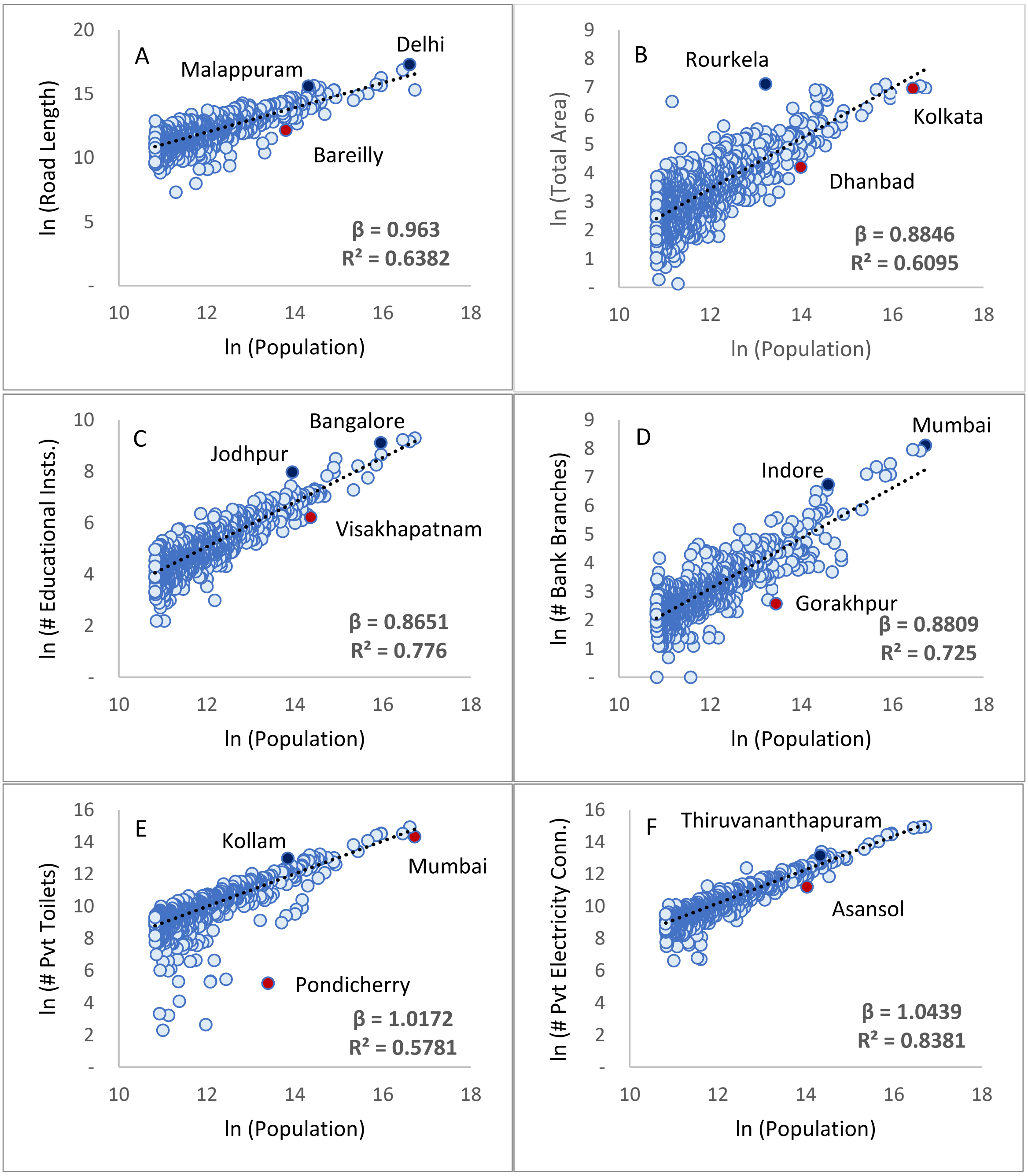

***Figure 1:*** **Scaling of public and private infrastructure with population.** Each panel shows the total value for each city (light blue circle) and the scaling best fit line, Eq. (1); the best fit estimate for the exponent $\beta$ and the goodness of fit $R^2$ are also shown, see Table 1 and Appendix A for details. Urban Indicators shown are A. Road Length, B. Total Area, C. Educational Institutions, D. Bank Branches, E. Private Toilets and F. Private Electricity Connections. All data from the Census of India 2011, see Appendix A for details.

## 3. GDP, Innovation and Crime in Indian Cities

We now turn to socioeconomic characteristics. First, consider the classic quantity measuring the size of the economy, the Gross Domestic Product (GDP). There are presently no official GDP statistics at the urban agglomeration level from the Government of India or state governments. The lowest level of spatial disaggregation for GDP data appears to be at the district level, which is not specifically urban: large cities span parts of several districts, whereas small cities are contained, together with peri-urban and rural areas, in other districts. Because this analysis is on different units of analysis and requires careful accounting of what is urban we defer it to future work.

This gap in such an important statistic has generated several estimates from non-official sources (details in Appendix A) – one from 2008 covering 13 cities measuring GDP in Purchasing Power Parity (PPP) terms and the other from 2010 covering 9 cities only, measuring nominal GDP, see Figure 2. These data sets are inconsistent with each other, one suggests a superlinear scaling with $\beta \simeq 1.12$ (roughly in line with other nations and theory), while the other suggests a sublinear relationship with $\beta \simeq 0.95$. Other data sources provide partial proxies for higher value economic activity in cities. For instance, the number of commercial and industrial electricity connections (not power usage) shows superlinear scaling with city size, with an exponent of $\beta \simeq 1.08$. In the absence of larger and more reliable datasets, it is difficult to say which of these relationships reflect the reality of GDP scaling, and the strength of economic agglomeration, in urban India.

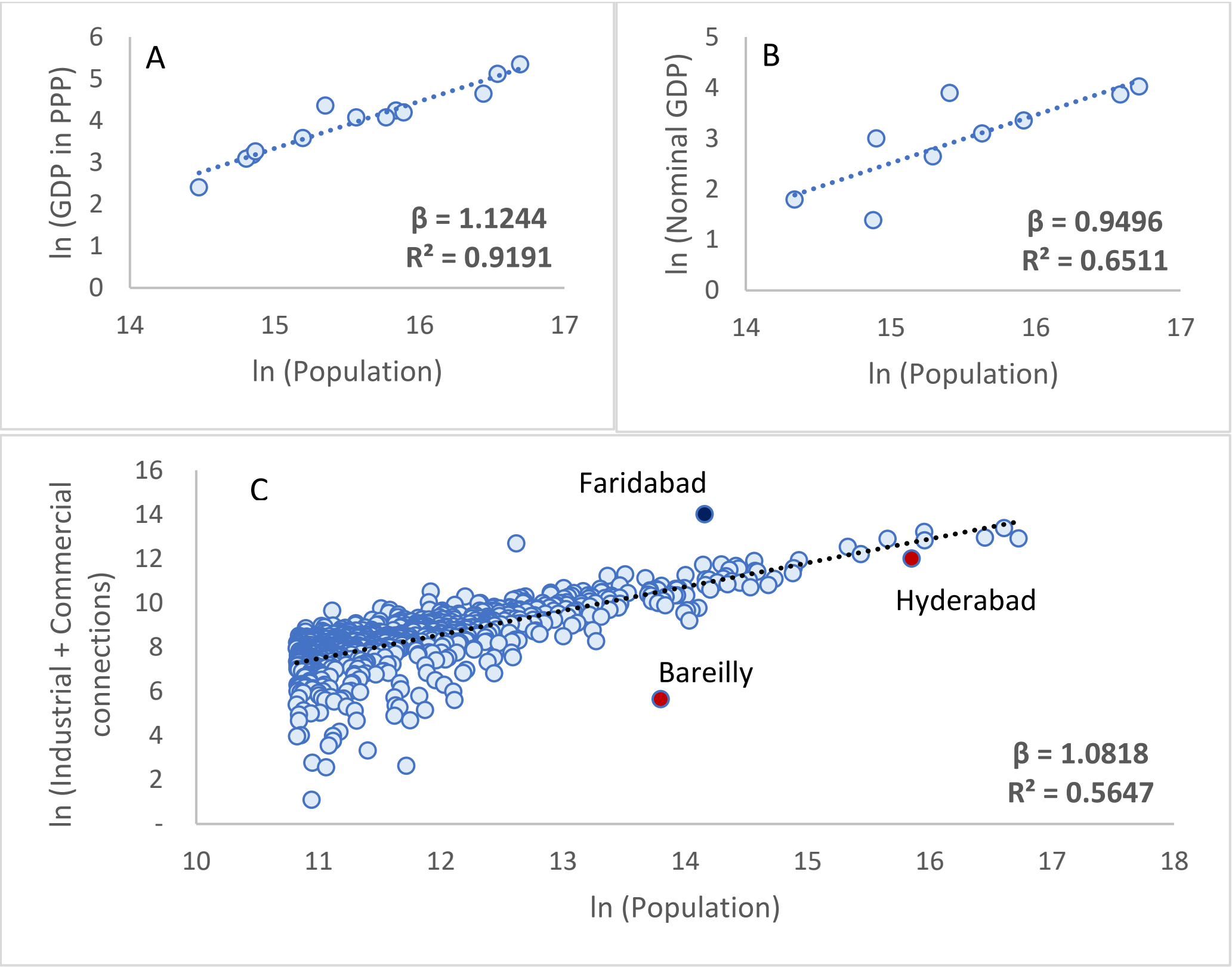

**Figure 2: Scaling of GDP with population from several partial sources and proxies**. A: GDP estimate with international Purchasing Power Parity for 13 cities estimated by PricewaterhouseCoopers (in USD). B: Nominal GDP (in USD) estimated for 9 cities by McKinsey (8 of which are also contained in the PwC dataset, but with vastly different estimates in many cases). C: GDP proxied by the total numbers of industrial and commercial electricity connections in Indian cities from the Census of India 2011.

We turn to assessing technological innovation rates in Indian cities. The classical quantitative measure for innovation is patent applications. However, the annual number of patents is zero for some cities in certain years. To overcome this issue, we use logarithmic binning of cities and data (see Appendix A) [28]. Figure 3A plots the scaling of logarithmic bins of patents with population and reveals a strongly superlinear relationship with $\beta \simeq 1.53$. Such a high $\beta$ is in keeping with observations on patent scaling from other jurisdictions (specifically the United States), where the scaling exponent is found to be significantly above 1, though below 1.5 [11,12]. Theoretical work using a network model of the city posits that superlinear scaling for innovation is a robust result (with a wide range of values between 1 and 2 for the exponent), obtainable using fairly loose assumptions [14,29].

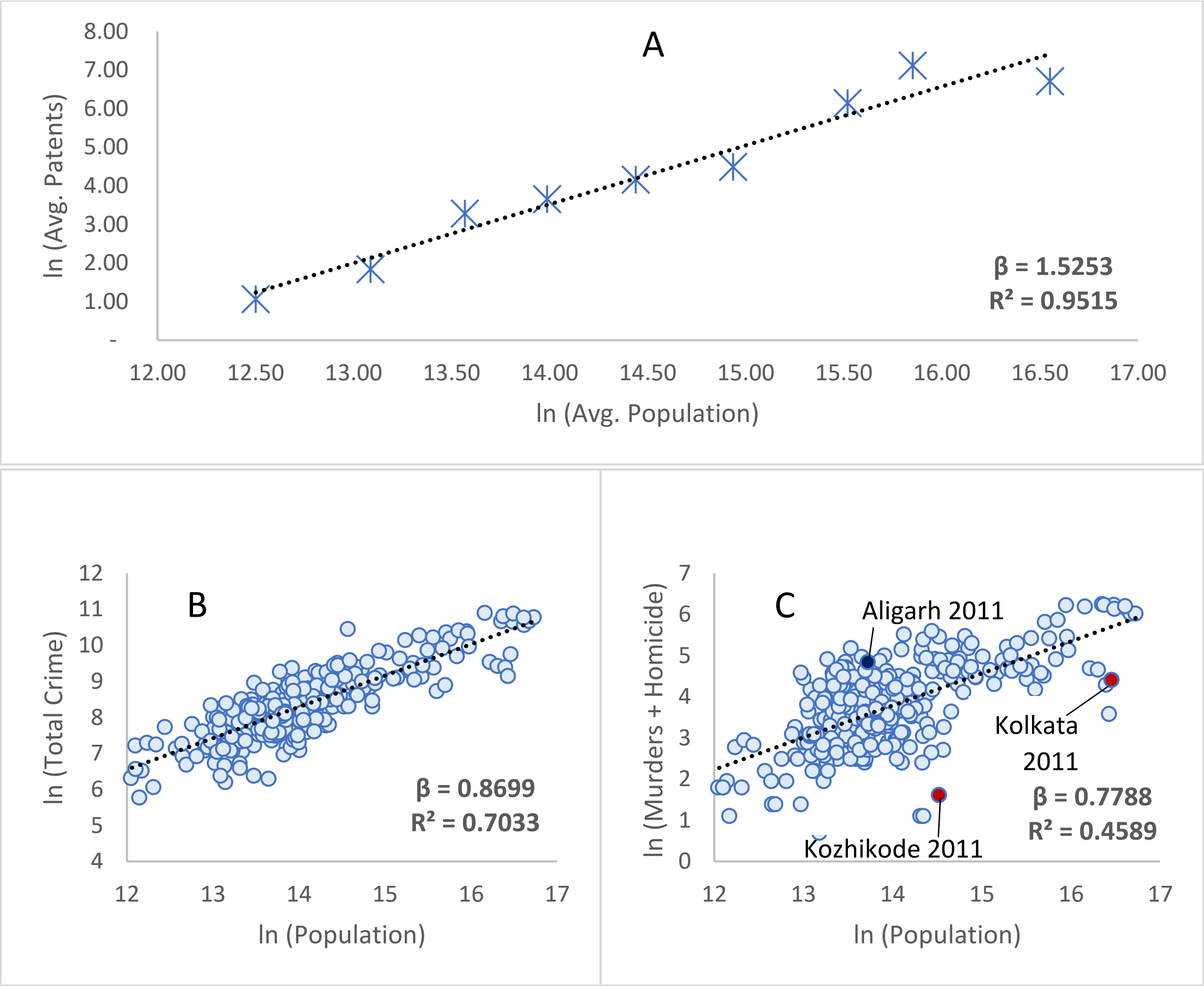

**Figure 3: Scaling of measures of innovation and crime with population**. A: Patents filled with the Office of the Controller General of Patents Designs and Trademarks (Intellectual Property India) for the years 2004, 2006, 2008, and 2011, averaged in logarithmic bins. B: Total crime, and C: Number of Murders and Culpable Homicides under the Indian Penal Code from the National Crime Records Bureau for the years 1991, 1996, 2001, 2006, and 2011.

Finally, we analyse crime rates in Indian cities. Figure 3B plots the scaling of total crimes (defined under the Indian Penal Code) with city population. We find a clear sublinear scaling relationship with population, with an exponent $\beta \simeq 0.87$. This result stands in stark contrast with the results for crime scaling in other contexts such as the United States [11] and several Latin American nations [16], where there is a superlinear relationship.

It has, however, been argued that total crime data in India in affected by significant under-reporting because of several social and structural factors [30], and that using this data to understand crime in India could yield an unrealistic picture. An observed sublinear scaling versus an expected superlinear trend (proportional to number of interactions) means, of course, that in large Indian cities socialization is substantially more peaceful than in smaller ones.

One subset of crime - serious crimes involving loss of life, such as murders and homicides - is posited to be a reasonably accurate representation of reality [16,30]; therefore, we attempt to understand the scaling of such serious crimes with population. Figure 3C plots the total numbers of murders and culpable homicides against population. This confirms the sublinear scaling relationship for crime, showing an estimated exponent that is in fact smaller than that for total crime ($\beta \simeq 0.78$).

**Table 1 summarises the specific statistics of scaling in the Indian context**.

| *Y* | *β* | *95% CI* | $R^2$ | *Number of Observations* | *Data* |
|---|---|---|---|---|---|
| Road length | 0.96 | 0.92 – 1.01 | 0.64 | 903 | Census 2011 |
| Total Area | 0.88 | 0.84 – 0.93 | 0.61 | 909 | Census 2011 |
| Number of educational institutions | 0.87 | 0.83 – 0.90 | 0.78 | 909 | Census 2011 |
| Number of bank branches | 0.88 | 0.85 – 0.92 | 0.72 | 908 | Census 2011 |
| | | | | | |
| Number of private toilets | 1.02 | 0.96 – 1.07 | 0.58 | 909 | Census 2011 |
| Number of private electricity connections | 1.04 | 1.01 – 1.07 | 0.84 | 909 | Census 2011 |
| | | | | | |
| Gross Domestic Product (PPP) | 1.12 | 0.90 – 1.35 | 0.92 | 13 | PwC 2008 |
| Gross Domestic Product (Nominal) | 0.95 | 0.33 – 1.57 | 0.65 | 9 | McKinsey 2010 |
| Number of commercial and industrial electricity connections | 1.08 | 1.02 – 1.14 | 0.56 | 903 | Census 2011 |
| Number of Patents | 1.53 | 1.40 – 1.70 | 0.95 | 320 | IPI 2004, 2006, 2008, 2011 |
| Number of crimes | 0.87 | 0.81 – 0.93 | 0.70 | 317 | NCRB 1991, 1996, 2001, 2006, 2011 |
| Number of serious crimes | 0.78 | 0.69 – 0.87 | 0.46 | 317 | NCRB 1991, 1996, 2001, 2006, 2011 |

In both the case of crime and innovation, scaling analysis provides only an approximate picture of urban dynamics. India is a continent-sized nation, with many regional differences, grounded in historic, cultural and political differences across the country. These differences lead to a characteristic urban geography of crime and innovation, which we analyse in the next sections.

## 4. The Urban Geography of Crime

Deviations from the average population size dependence (scaling) estimated in the previous section, give us a principled way to assess and characterize these local and regional effects [17,31,32]. The residuals from scaling (the vertical distance of each point from the fit line in Figs 1 and 2) gives us a *Scale Adjusted Metropolitan Indicator* (SAMI), $\xi_i(t)$, which characterizes city *i* at time *t:*

$$\xi_i(t) = \log \frac{Y_i(t)}{Y_0(t) N_i^{\beta}} \qquad (4)$$

This is a dimensionless metric that makes direct comparison between the performance of different cities possible because population size agglomeration effects have been factored out [31].

We now calculate and analyse the SAMIs for patents and crime in India. This will allow us to better assess the nature of local dynamics and also, in the case of crime, potentially explain the counter-intuitive sublinear scaling effect.

Figure 4A plots the SAMIs for crime in Indian cities in rank order from more crime than expected to less. Aligarh (Uttar Pradesh) is ranked 1 (worst), while Malappuram (Kerala) is the best performing city in India. To illustrate the difference between the standard per capita measure of crime and crime SAMIs (which removes population size bias) we show the expectations from these two approaches for the 10 largest Indian cities. Each one of these cities performs worse when ranked by SAMIs than when measured by a standard crime per capita measure (Figure 4B). For instance, the per capita crime measure shows us that only 1 of these 10 cities is in the top half of cities ranked in descending order of crime, while the corresponding SAMI analysis shows that 5 out of these 10 cities are in the top half. This is because, crime rates being sublinear in population, the SAMIs contain the expectation that larger cities should have less crime per capita and measures only the deviation from such expectation.

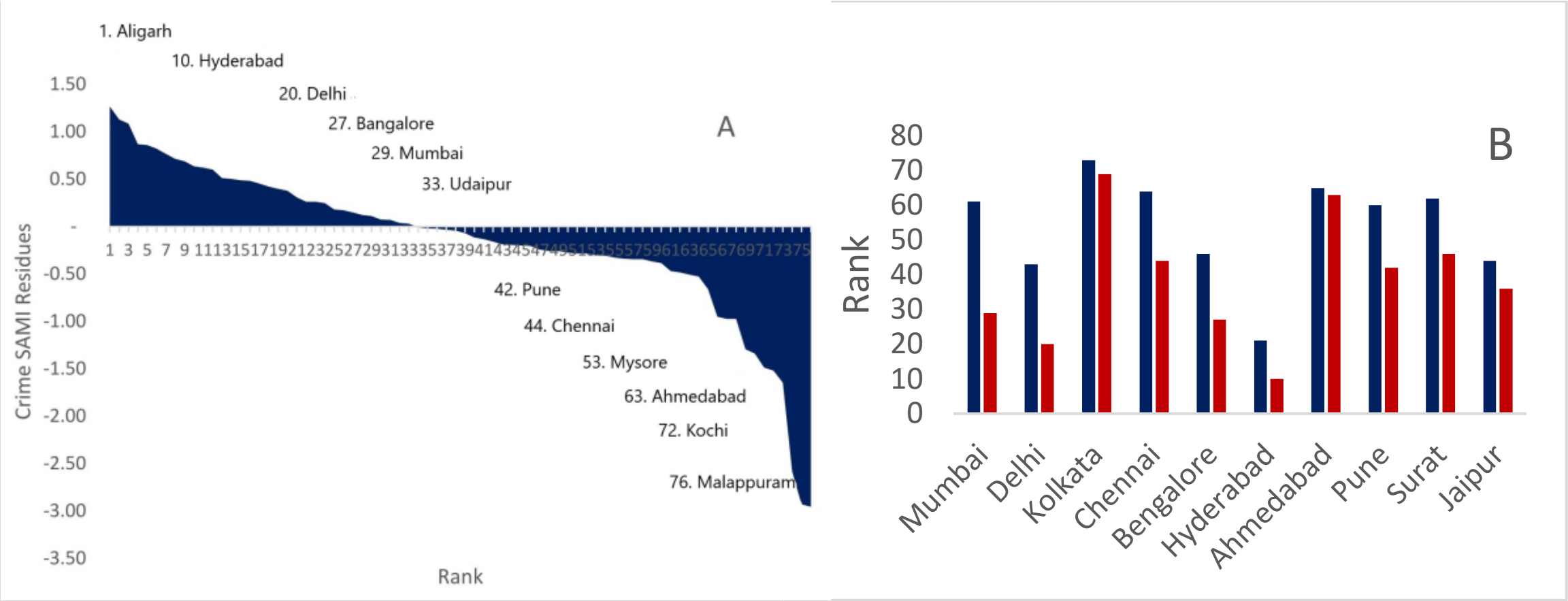

***Figure 4:*** **Analysing residuals and per capita metrics for crime**. A: Rank order of Crime SAMI Residues 2011 B: Crime Rank 2011 by City. Red: Crime SAMI residue rank. Blue: Crime per capita rank.

Crime in India is arguably an under-researched subject. Studies that explore the theme as a sociological phenomenon do so through the lens of gender [30,33] and caste [34]. The caste system in India today emerges from the ancient ‘varna’ system, where society was divided into five hereditary, endogamous, and occupation-specific groups [34]. The lowest castes and the Dalits have historically suffered violence at the hands of the upper castes, but post-independence governments in India have attempted to counter this history by expanding the scope of affirmative action and passing specific legislation on crimes against Dalits [34]. While these interventions have resulted in improvements on public goods provision [35] and redistribution of resources [36] to disadvantaged groups, social and political organization among Dalits has often resulted in feudal backlashes taking the form of “mass killings, gang rapes, looting in Dalit villages” [37]. Analysis of data from the National Council for Scheduled Castes (NCSC) shows that 60% of the atrocities committed against Dalits in India occurs in four states – Uttar Pradesh, Rajasthan, Bihar, and Madhya Pradesh [37]. This data also reveals that most serious crimes against Dalits including hate crimes, rape, and murder occur predominantly in rural settings or in small urban settlements.

Gender is the other sociological lens applied to understand crime in India. Dreze and Khera [30] find a robust negative correlation between female to male ratio and murder rates. While causality is hard to assess, they argue that low female-male ratios and high murder rates are both manifestations of patriarchy. This resonates also with the role of evolutionary psychology in the incidence of crime [38]. Given that patriarchal subjugation of women is based on violence (or its implied threat), Dreze and Khera [30] posit that we would indeed expect that areas with high violence would be associated with sharp gender inequalities. This result is also in agreement with Oldenburg [33] on district level data for the northern Indian state of Uttar Pradesh, where he found negative correlation between incidence of murders and female-male ratio. He hypothesised a causal influence of violence on female-male ratios, arguing that in regions of high violence, male child preference would be particularly high as sons are seen both as protection against violence and as potential candidates to exercise power [33]. It is however at the intersection of gender with caste that violent crime becomes salient - in the form of 'honour crimes'. While ostensibly the legal system in India is free from the grip of caste, there exists a parallel institution of caste-based panchayats (local level clan assemblies) that function based on their own traditional law [39]. Caste panchayats actively rule against inter-caste marriages and even proclaim death as punishment in many cases, also called 'honour' killings [39]. While 'honour' killings are a pan-Indian phenomenon, it is especially prevalent in the states of northern India – specifically Uttar Pradesh, Delhi, Rajasthan, and Haryana [40].

We now assess the geographic spread of crime SAMIs to investigate if there is support for these contentions. Figure 5A maps the spread of crime SAMIs in India and we can clearly see the distinct geographical tilt in crime behaviour in India. Even a cursory visual inspection makes it clear that a plurality of cities in north and central India, large and small, have high crime SAMIs, while those in western and southern India have much lower crime SAMIs. The higher crime SAMIs in the smaller cities of north-central India, especially given the salience of caste in generating crime, is possibly correlated to the fact that smaller settlements have a higher proportion of Dalit population. We find evidence for higher Dalit populations in smaller cities when we examine the scaling of Dalit population with city size and discover sublinear scaling with an exponent of 0.96.

We also split the scaling plot for crime into two groups of cities – one group for cities in North and Central India and another group for Southern and Western cities. Figure 5B shows that the set of Northern-central Indian cities has a significantly lower $\beta$ than the southern-western cities group (0.63 compared to 0.92) because small and medium cities in the northern-central group have higher crime SAMIs than comparable cities in the southern-western set. It is also important to note that even though there is a significant difference in $\beta$ between the two groups, both are still sublinear. This is potentially a reflection of the fact that despite significant differences in crime behaviour across geographies in India, there is indeed an underlying sociology of culture-specific crime (honour and caste-based crime) that is at work across the nation.

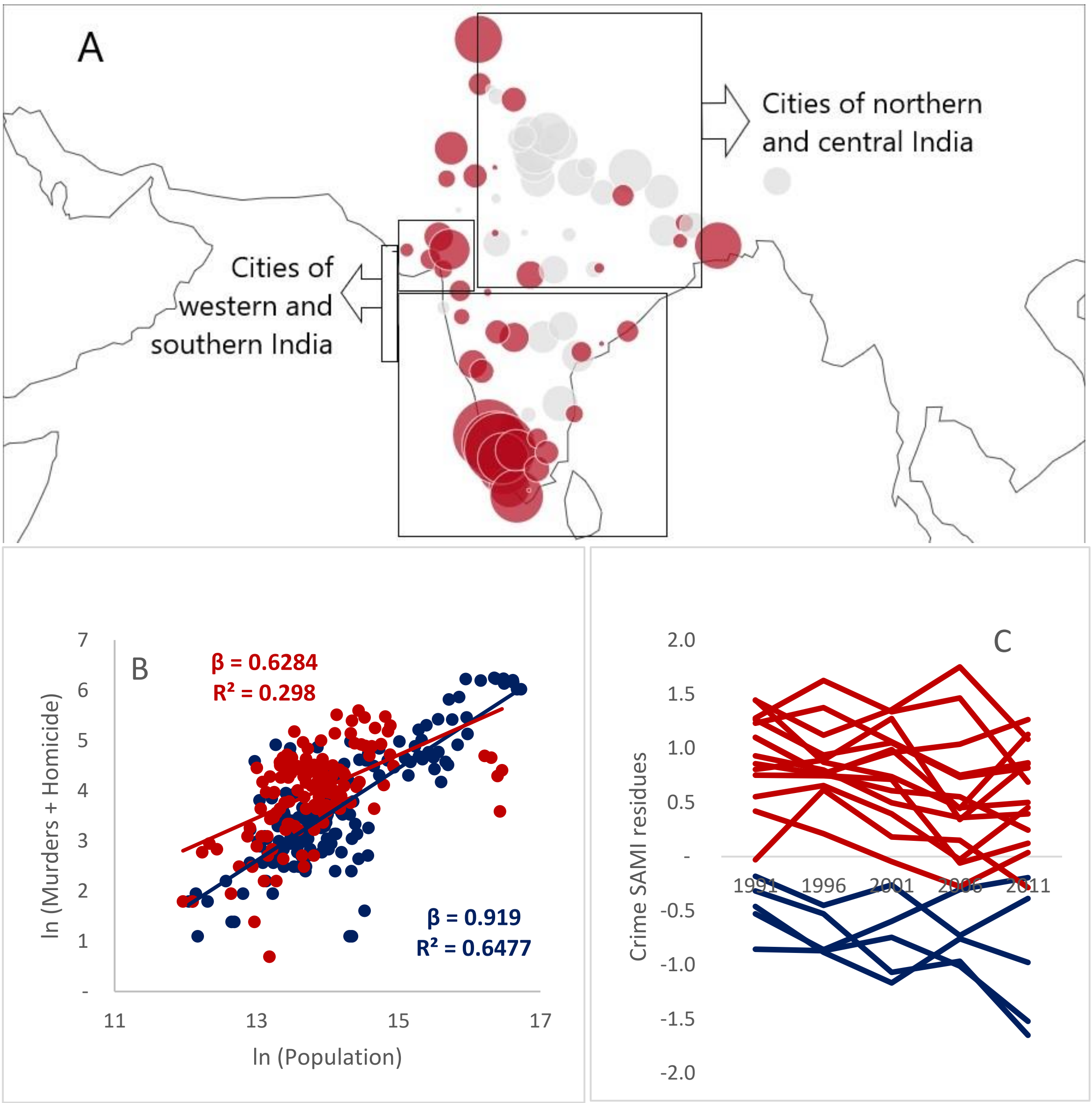

***Figure 5:*** **Spatial and temporal analysis of crime residuals**. A: Spatial distribution of Crime SAMIs in 2011: Red (grey) dots correspond to deviations below (above) expectation for city size. The size of the circle denotes the magnitude of Crime SAMIs. B: Murders & culpable homicide vs. population. Red group: Cities in northern and central India. Blue group: Cities in Southern and Western India. C: Temporal evolution of crime SAMIs for select cities between 1991 and 2011. Blue lines: Cities in northern and central Indian states of Uttar Pradesh (Agra, Aligarh, Allahabad, Bareilly, Gorakhpur, Kanpur, Lucknow, Meerut, Moradabad, Varanasi), Madhya Pradesh (Bhopal, Indore, Jabalpur), and Bihar (Patna). Red lines: Cities in southern Indian states of Kerala (Kochi, Thiruvananthapuram), and Tamil Nadu (Chennai, Coimbatore, Madurai).

These geographical differences also appear to be contingent on history, as evinced in Figure 5C, which plots the temporal evolution for crime SAMIs for cities in five Indian states between 1991 and 2011. The temporal evolution of crime SAMIs in Uttar Pradesh, Madhya Pradesh, and Bihar show that deviations from scaling have remained high throughout this period, while the temporal pattern of crime SAMIs for cities in the southern states of Kerala and Tamil Nadu have remained low. This is an indication that even as cities gain population over decades, local characteristics can persist over long time periods [31].

We now seek to validate these patterns more formally by attempting to cluster cities based on the distance between their SAMIs. Figure 6 plots a heatmap of the Euclidean distance between pairs of SAMIs. We see eight clusters of cities, with five clusters comprised largely of northern and central Indian cities and the other three clusters of southern and western Indian cities. This decomposition provides a formal confirmation of the spatial spread of SAMIs in Figure 5A and suggests that regional variations in caste and gender dynamics could be critical to the nature of homicides and murders observed in India.

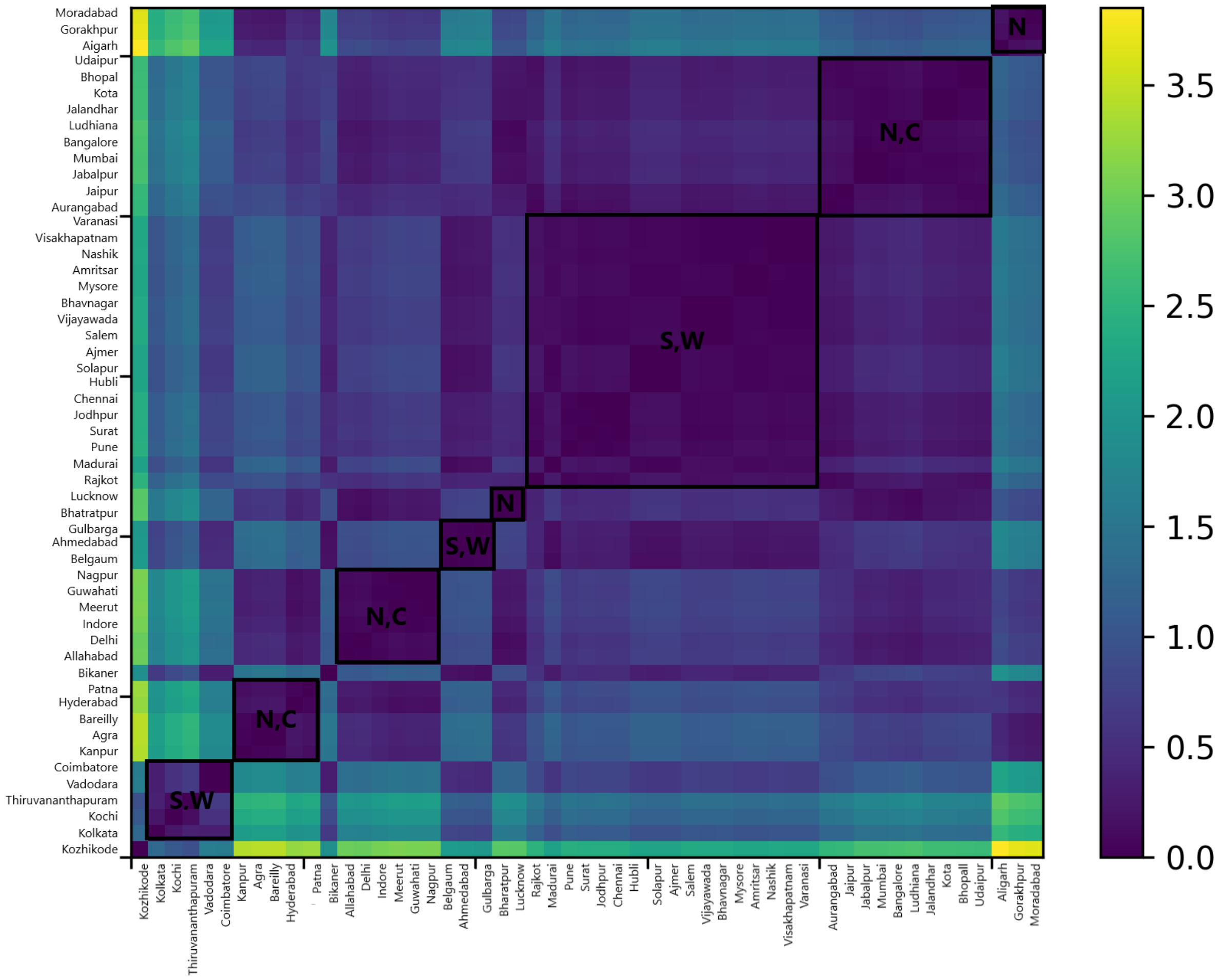

**Figure 6: Heatmap of Crime SAMIs for Indian cities (2011).** Clusters of cities are based on the Euclidean distance between SAMIs (darker blue indicate smaller distances**).** Eight clusters of cities are demarcated and represented by the regions they belong primarily to: N: Northern India, C: Central India, S: Southern India, and W: Western India.

We also assess how spatial distance affects crime SAMI behaviour. Spatial similarity between cities $i$ and $j$, $c_{ij}$, is computed as the equal-time cross-correlation of their SAMI time series [31]:

$$c_{ij} = \frac{1}{|\xi_i||\xi_j|} \sum_t \xi_i(t)\xi_j(t) \qquad (5)$$

This measure ensures that cities with similar SAMIs and time series have high correlation. Figure 7 plots cross-correlation as a function of distance between cities. We find that while short distance correlation appears to exist for up to approximately 300 km, this effect disappears for greater distances.

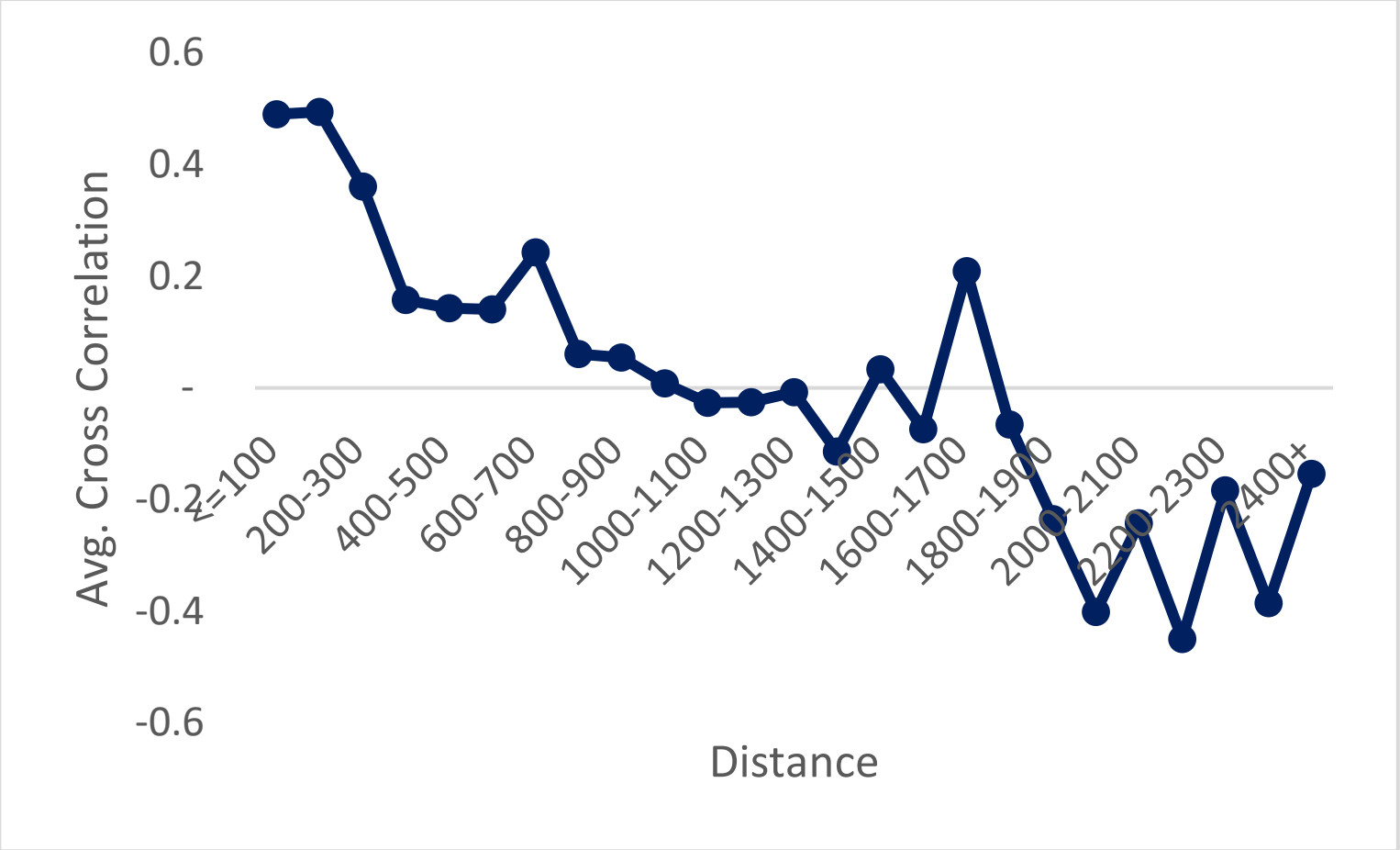

***Figure 7:*** **Cross-correlation of crime residues and distance between cities.** Average cross-correlation of Crime SAMIs of cities (1991 – 2011) vs. Straight line distance between cities

To conclude our analysis of crime in Indian cities, we provide some international comparisons to put the Indian evidence into a broader context. UNODC data reveals that while the global intentional homicide rate is 6.2 (per 100,000 inhabitants), there are significant regional variations, with the Americas and Africa exhibiting high rates of 16.3 and 12.5, and Europe, Oceania, and Asia showing much lower rates of around 3 [41]. Within these regional classifications, there are also significant national variations. The intentional homicide rate in India is found to be 3.2, broadly in line with the Asian average, and lower that of other large countries like Brazil (29.5), Nigeria (9.9), or the United States (5.4), but considerably higher than China (0.6) or Indonesia (0.5). Within nations, when we compare the rates of crime in Indian and American cities, we find that the largest Urban Agglomerations in India – Mumbai, Delhi, and Kolkata – have homicide (murder and culpable homicide) rates of 2.2, 3.0, and 0.6 (see Appendix A), while the corresponding homicide rates in the largest metropolitan areas in the United States – New York, Los Angeles, and Chicago - are 3.3, 4.9, and 7.1 respectively [42]. Therefore, serious crime in India appears to broadly conform within the regional (Asian) benchmarks, while homicide rates in large Indian cities are significantly lower than their counterparts in the United States.

## 4. The Urban Geography of Technological Innovation in India

We now turn to an analysis of the deviations in innovation SAMIs, Figure 8A. Again, we find significant discrepancies between the rank ordering of cities based on innovation SAMIs and innovation measured as patents per capita (Figure 8B). When ordered by patents per capita, 5 of the 10 largest Indian cities are ranked among the top 10 most innovative cities, but when ranked by innovation SAMI, only one of them, Bangalore, appears in the top 10. This finding points to many smaller and medium cities that are quite inventive for their population size and that should be the focus of some additional attention in terms of both scholarship and policy.

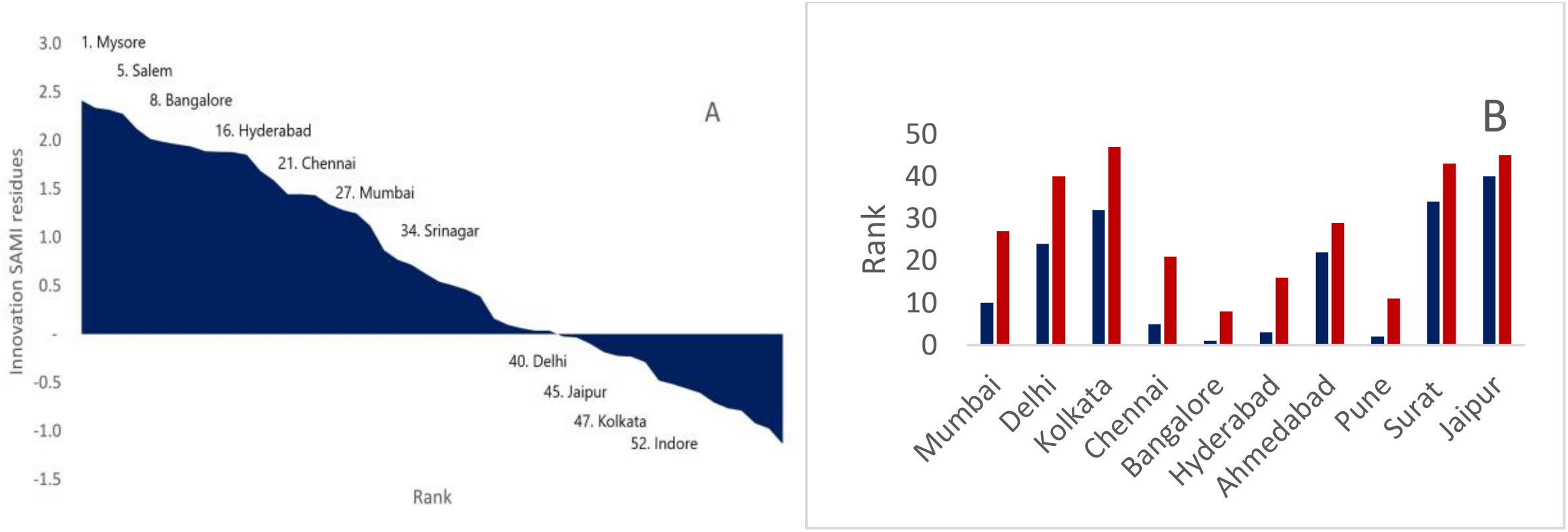

**Figure 8: Analysing residuals and per capita metrics for innovation**. A: Rank order of Innovation SAMI residuals in 2011 B: Innovation Rank 2011 by City. Red: Innovation SAMI residuals rank. Blue: Innovation per capita rank.

Figure 9 reveals that the ten most innovative cities in India, once one accounts for strong population size scaling, are in fact small and medium cities with a median population of 1.02 million and that their innovation SAMIs have remained consistently high in the period 2006-2011, suggesting temporal persistence of innovation. For instance, cities like Jamshedpur, Trichy, Salem, and Ranchi, have historically been centres of heavy industry, while Bangalore and Mysore are centres of Information Technology. A deeper analysis is required to understand sectors and drivers of innovation across the relatively smaller urban centres.

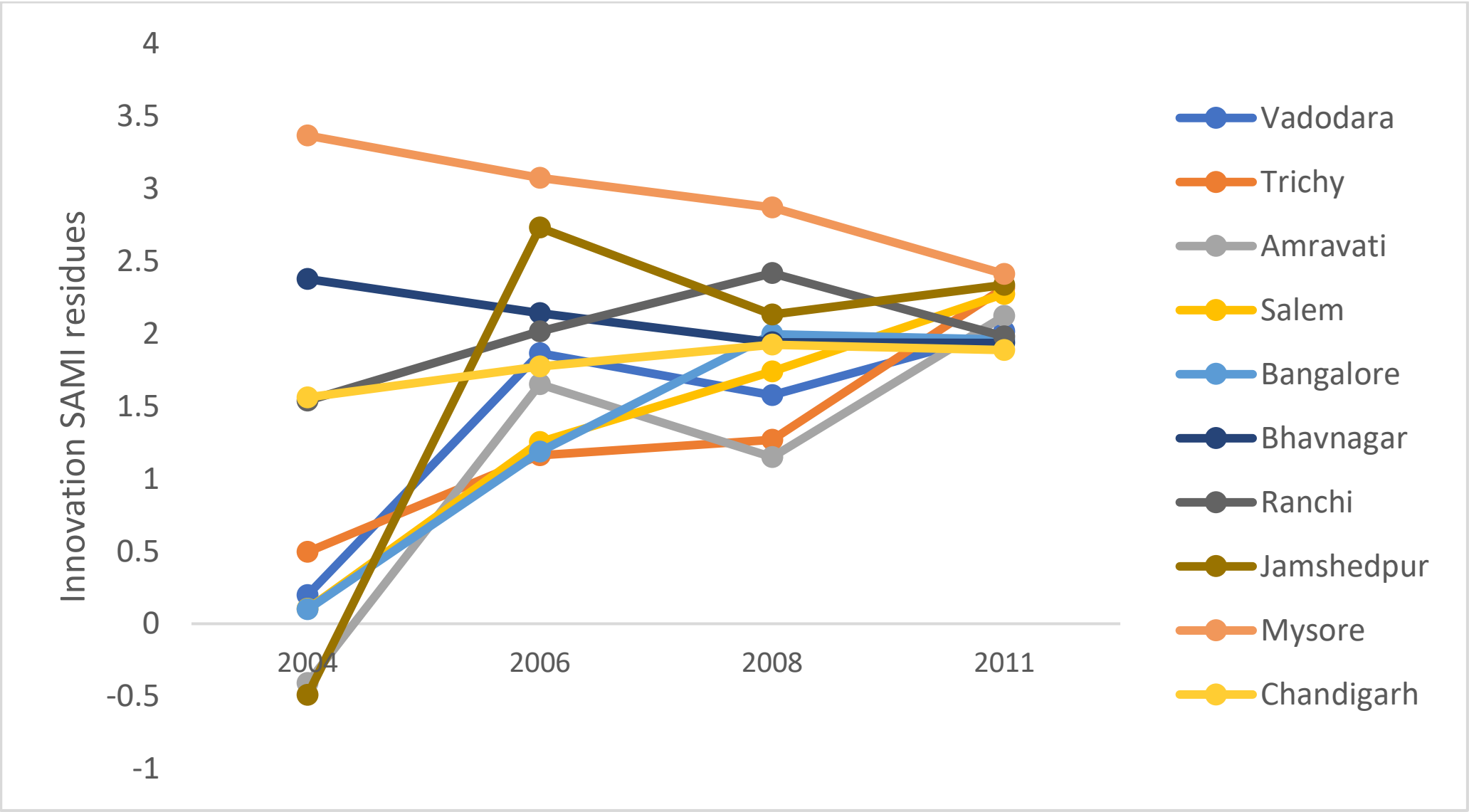

**Figure 9: Temporal analysis of innovation scaling residuals (2004 – 2011) for top 10 cities ranked by SAMIs for patents.**

We also map the innovation SAMIs across India in Figure 10 and find that spatially, cities in the South and West of India appear generally more likely to be innovation hotspots, while cities in the North, Centre, and East of the country appear to be lagging. However, it is important to point out that while this seems to follow the same overall pattern of crime SAMIs, the extent of regional bias in this case does not appear to be as strong.

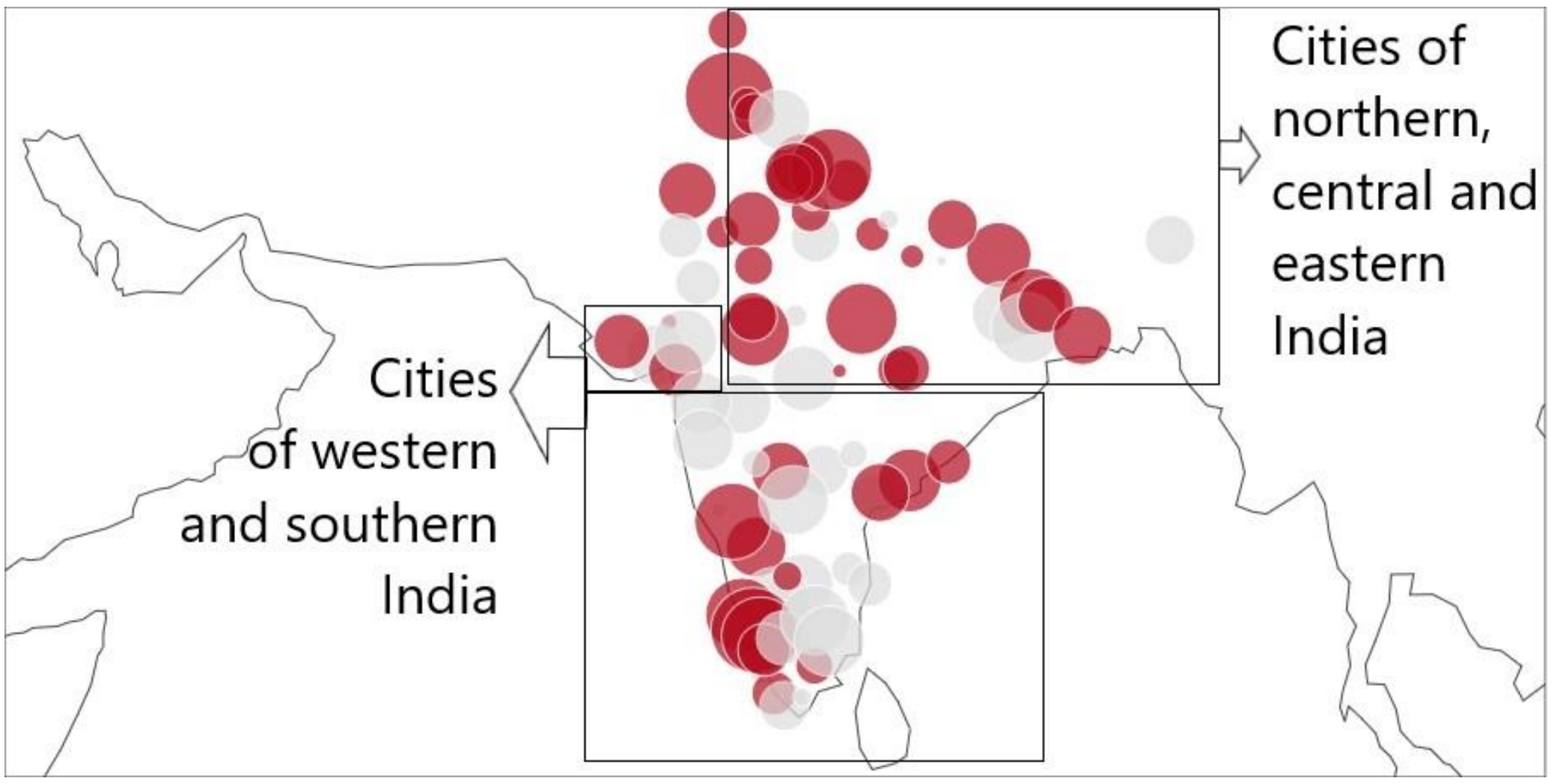

***Figure 10:*** **Spatial analysis of innovation SAMI residuals in 2011.** Red (grey) dots correspond to deviations above (below) expectation for city size. The size of the circle denotes the magnitude of the corresponding Innovation SAMI.

As before, we try to confirm this intuition by clustering cities based on the Euclidean distance between their innovation SAMI. Figure 11 plots the heatmap of city clusters, and as is apparent, while there are indeed three clusters comprised largely of southern and western cities, and three clusters of northern, central, and eastern cities, there are also two significantly large geographically mixed clusters of cities. So, while there appears to be some extent of regional variation, the spatial relationship is not as stark as it was in the case of crime.

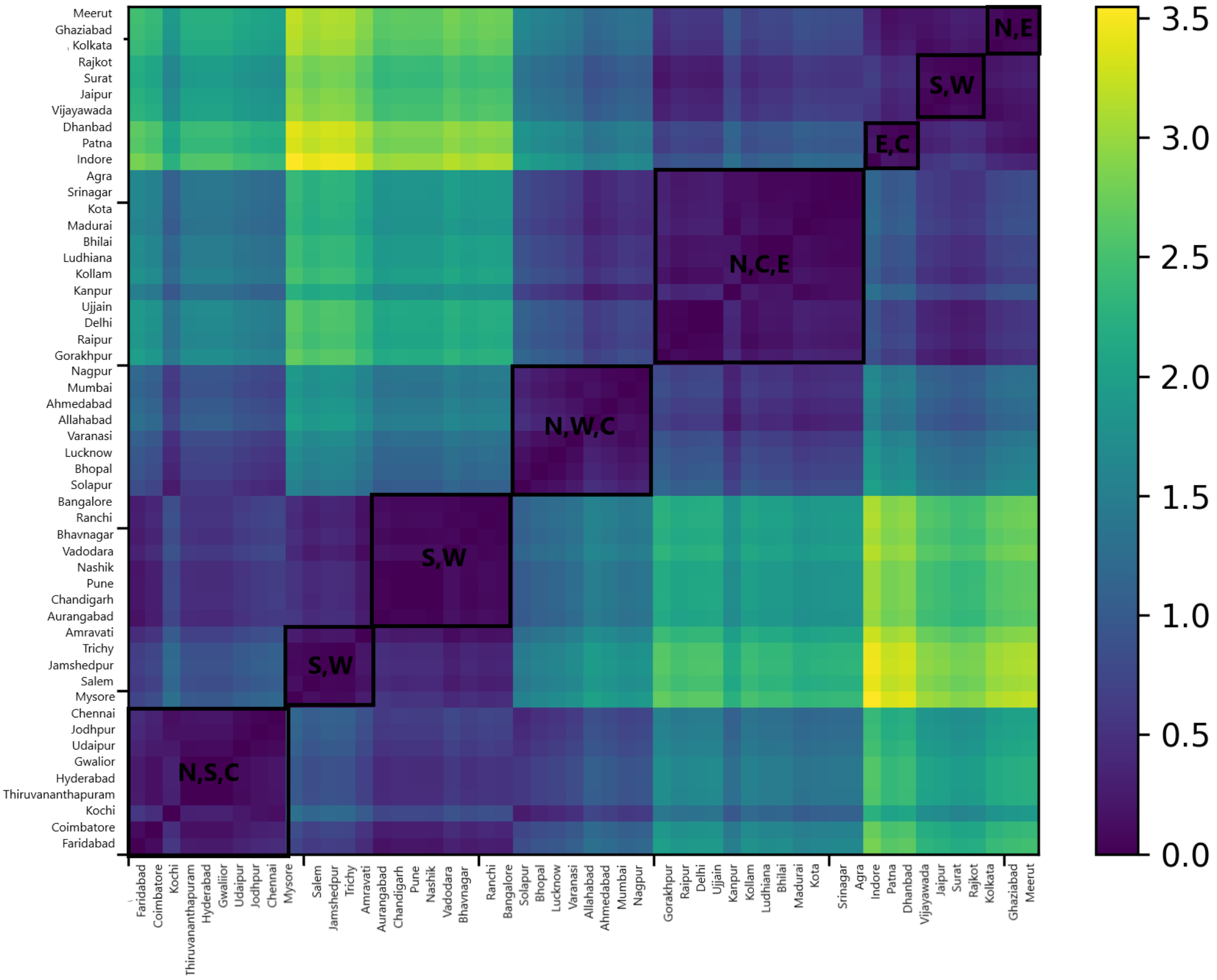

**Figure 11: Heatmap of Innovation SAMIs of Indian cities (2011)**. City clusters are built based on the Euclidean distance between SAMIs (darker blues indicate lesser distances). Eight clusters of cities are demarcated: Clusters are represented by N: Northern Indian cities, C: Central Indian cities, S: Southern Indian cities, E: Eastern Indian cities and W: Western Indian cities.

## 4. Conclusions

We have analysed emerging data for Indian cities in light of urban theory and known statistical patterns for other urban systems, with the aim of characterizing the tell-tale signals of urbanization in terms of scaling and agglomeration economies, and geographic patterns of development and variation.

Although empirical information at the level of cities is becoming more available in India from official sources, such as the Census of India (2011), the paucity of data for functional urban definitions relative to most other high- and middle-income countries makes our analysis necessarily limited and tentative. Within these limitations, we find patterns of urban density scaling with population size roughly in line with other urban systems and historical cases. Regarding infrastructure delivery, large Indian cities seem to have an advantage relative to smaller towns, which is a pattern typical of other urban systems where basic infrastructure such as roads, sanitation and electricity access are not yet universal and spread from larger urban areas to other parts of the country [15]. One of the most critical gaps in India is the ability to assess the size and development of urban economies. We discussed existing data and pointed to some existing contradictions that need to be resolved in order to understand and harness the potential of Indian cities for economic growth. Innovation, measured by patenting activity, shows a strong superlinear pattern with city size (similar to other nations, such as the United States), meaning that it is disproportionally concentrated in larger urban areas. Within this general pattern, however, we were able to identify many smaller cities with uncharacteristically high patent productivity and discuss India's detailed geography of technological innovation.

Arguably, the biggest surprise about Indian cities is the sublinear pattern of crime, including murders and homicides, which – unlike in most developed nations – translates in higher rates of violence per capita in smaller towns, relative to the nation's largest cities. Though many questions about the data remain, we were able to derive the geography of crime across India and relate it to more specific studies that identify most sources of violence in the country associated with issues of gender and caste. These show a strong regional signature and have been discussed by sociologists and anthropologists as a predominantly rural or small city phenomenon.

Indian urbanization, currently estimated at 33%, is expected to rise to 53% by 2050 [2], adding hundreds of millions of people to cities and creating giant megacities, with perhaps as many as 50 million people over that period. While we hope that this paper presents the beginning of a holistic empirical characterization of Indian cities, there is a critical need for a concerted effort aimed at measuring urban economic statistics at the local level, including in neighborhoods, which tend to express the strongest patterns of concentrated (dis)advantage and thus inequality [15]. Successful Indian urbanization is critical not only for well-being of all people in India, but for the sustainability of the entire planet. We cannot afford to fly blind through this momentous transformation.

## Appendix A: Data sources and methods

*Population data*: City level population is available from the Census of India at http://www.censusindia.gov.in/2011census/dchb/DCHB.html. To consolidate city level population into population at the Urban Agglomeration (UA) level, the composition of 298 UAs in India is again available from the Census of India data, tabulated and provided at http://www.census2011.co.in/urbanagglomeration.php. An *Urban Agglomeration* is defined by the Census of India as ".. a continuous urban spread constituting a town and its adjoining urban outgrowths (OGs) or two or more physically contiguous towns together and any adjoining urban outgrowths of such towns.”. An urban outgrowth (OG) is defined as “.. a viable unit .. contiguous to a statutory town .. possess(ing) urban features in terms of infrastructure and amenities ..”. A complete list of census concepts and definitions is available at http://censusindia.gov.in/2011-prov-results/paper2/data_files/kerala/13-concept-34.pdf. For our analysis, we consider all *urban agglomerations* with populations of at least 50,000.

*Infrastructure data*: Data on city level infrastructure is available from the Census of India at http://www.censusindia.gov.in/2011census/dchb/DCHB.html, under the column “Town Amenities”. Each file contains data for one state. This data is available (for each state) at the city level and conversion to UA level data is just as described for population data. This has to be done for every state. For our scaling analysis, we build the infrastructure metrics from the raw data provided by the census in the following manner:

1. Road length = Pucca Road Length + Kuccha Road Length
2. Number of educational institutions = Schools (including primary, middle, secondary and senior secondary, both government and private) + Colleges (including arts, science, commerce, arts and science, arts and commerce, arts science and commerce, law, university, medical, engineering, management, and others, both government and private) + Polytechnics (government and private)
3. Number of bank branches = Nationalised bank branches + Private bank branches + Cooperative bank branches
4. Number of private toilets is available as latrine count
5. Number of private electricity collections is also directly provided in raw data
6. Number of commercial and industrial electricity connections = Industrial connections + Commercial connections + Other connections
7. Total Area is directly available in the raw data

For all infrastructure, public and private, we use data from the 2011 census for the scaling analysis. The complete data set has 911 data points.

*Gross Domestic Product (GDP) data*: There is no official data series of urban GDP in India, so we looked for other sources for this data. We found two small datasets:

1. Price Waterhouse Coopers’ 2009 UK Economic Outlook report lists GDP data for 13 Indian cities for 2008. The complete list is available at: https://en.wikipedia.org/wiki/List_of_cities_by_GDP. This data is not drawn from any official series of the Government of India but estimated by PWC. The methodology and approach to estimating GDP is available in Annex B of the report: https://web.archive.org/web/20110504031739/https://www.ukmediacentre.pwc.com/imagelibrary/downloadMedia.ashx?MediaDetailsID=1562. The report estimates GDP at Purchasing Power Parity (PPP) exchange rates to correct for price level differences between countries.
2. McKinsey estimated GDP for 9 Indian cities in 2010. The complete list is available at: https://en.wikipedia.org/wiki/List_of_cities_by_GDP. A sample of this list was published by Foreign Policy titled “The most dynamic cities of 2025”, a list of 75 cities around the world. This list contained 2010 GDP estimates for 3 Indian cities and is available at: https://foreignpolicy.com/2012/08/07/the-most-dynamic-cities-of-2025/. This estimate was for nominal GDP and did not incorporate PPP correction.

*Crime data*: Crime data is put out annually at city level by the National Crime Records Bureau (NCRB), and these reports titled “Crime in India” are available at http://ncrb.gov.in/. Crime is broken down into multiple heads. For our analysis:

1. Total crime = Total cognizable offences under the Indian Penal Code (IPC)

2. Murders and Homicide = Murder (Sec. 302 IPC) + Culpable Homicide not amounting to murder (Sec. 304 & 308 IPC)

We use crime data for the years 1991, 1996, 2001, 2006, and 2011. While direct census data is used for population numbers in 1991, 2001, and 2011, we interpolate the population numbers for 1996 and 2006 using the Compound Annual Growth Rate (CAGR) calculation for the periods 1991-2001 and 2001-2011 respectively. The complete data set has 317 data points.

*Innovation data*: For this data, we used the published patent search of Intellectual Property India at http://ipindiaservices.gov.in/publicsearch. Given that the data itself is not readily available in a document format, we had to individually search for data on each city analysed. We collect data for the years 2004, 2006, 2008, and 2011. However, given the fact that there were a few zero data points (cities where no patents were published for a given year), the scaling analysis is not directly performed on all the raw data points. We bin all the data in logarithmic bins (logbins) of population. For instance, the first logbin of population we use is 12.25-12.75, which is to say that for all cities whose *log(population)* is between 12.25 and 12.75, their patent counts are averaged, and the logarithm of this average patent count is taken. We therefore end up with 9 logbins of population from 12.25-12.75 to 16.25-16.75 and each of these logbins have a corresponding *log(average patent count)* measure. We plot the scaling relationship between these two derived values to arrive at the scaling exponent for innovation. The complete data set has 320 data points.